\documentclass[prb,showpacs,eqsecnum]{revtex4}

\usepackage{graphicx}
\usepackage{amsmath}
\usepackage{amssymb}

\newcommand{\be}{\begin{equation}}
\newcommand{\ee}{\end{equation}}
\newcommand{\ba}{\begin{eqnarray}}
\newcommand{\ea}{\end{eqnarray}}
\newcommand{\Si}{\mbox{Si}}

\newcount\bozza \bozza=0
\ifnum\bozza=1
\newdimen\shift \shift=-2truecm
\def\lb#1{%
{\label{#1}\rlap{\kern\shift{$\scriptstyle#1$}}}}
\else\def\lb#1{\label{#1}} \fi

\begin{document}

%\preprint{HEP/123-qed}

\title{Aharonov-Bohm effect in relativistic and nonrelativistic 2D electron gas:
a comparative study}

\author{A.O. Slobodeniuk}
%\thanks{On leave of absence from }
\email{aslobodeniuk@gmail.com}
\affiliation{Bogolyubov Institute for Theoretical Physics, 14-b
        Metrologicheskaya Street, Kiev, 03680, Ukraine}

\author{S.G.~Sharapov}
\email{sharapov@bitp.kiev.ua}
\affiliation{Bogolyubov Institute for Theoretical Physics, 14-b
        Metrologicheskaya Street, Kiev, 03680, Ukraine}

%\thanks{Present address: Bogolyubov Institute for Theoretical Physics, Kiev, Ukraine}
%\homepage{http://}

\author{V.M.~Loktev}
\email{vloktev@bitp.kiev.ua}
\affiliation{Bogolyubov Institute for Theoretical Physics, 14-b
        Metrologicheskaya Street, Kiev, 03680, Ukraine}
\affiliation{National Technical University of Ukraine "KPI",
   37 Peremogy Ave., Kiev  03056, Ukraine}

\date{\today }

\begin{abstract}
We carry out a comparative study of electronic
properties of 2D electron gas (2DEG) in a magnetic field of
an infinitesimally thin solenoid with relativistic dispersion as in graphene
and quadratic dispersion as in semiconducting heterostructures.
The problem of ambiguity of the zero mode solutions of the Dirac equation is treated  by
considering of a finite radius flux tube which  allows to select unique solutions associated
with each $\mathbf{K}$ point of graphene's Brillouin zone. Then this radius is allowed to go
to zero. On the base of the obtained in this case analytical solutions in the Aharonov-Bohm potential
the local and total density of states (DOS) are calculated.
It is shown that in the case of graphene there is an excess of LDOS near the vortex,
while in 2DEG the LDOS is depleted. This results in excess of the induced by the vortex
DOS in  graphene  and in its depletion in 2DEG. We discuss the application of the results for
the local density of states for the scanning tunneling spectroscopy  done on graphene.
\end{abstract}

%03.65.-w 	Quantum mechanics [see also 03.67.-a Quantum information; 05.30.-d Quantum statistical mechanics; %31.30.J- Relativistic and quantum electrodynamics (QED) effects in atoms, molecules, and ions in atomic physics]
% 81.05.Uw Carbon, diamond, graphite
%73.20.At 	Surface states, band structure, electron density of states
%72.10.Fk 	Scattering by point defects, dislocations, surfaces, and other imperfections (including Kondo effect) %73.50.-h 	Electronic transport phenomena in thin films (for electronic transport in mesoscopic systems, see %73.23.-b; see also 73.40.-c Electronic transport in interface structures; for electronic transport in nanoscale %materials and structures, see 73.63.-b)

%\keywords{QUANTUM HALL EFFECT; GRAPHENE}

\maketitle

\section{Introduction}
\label{sec:intro}

Physics of graphene perception begins when one compares Landau levels in two dimensional Schr\"{o}dinger and
Dirac theories. Such spectacular phenomenon as unconventional quantum Hall effect
\cite{Geim2005Nature,Kim2005Nature} is caused by the anomaly of the lowest Landau level (LLL) \cite{Gusynin2005PRL}
which for Dirac fermions in graphene is field independent and can accommodate only half the usual number
of the states from the conduction band and takes the other half from the valence band.
The easiest way to accomplish this peculiar feature of the LLL  is to solve a pair of the Dirac equations
that describe excitations near two inequivalent $\mathbf{K}$ points of graphene's
Brillouin zone. Normally this is done in a constant  homogeneous magnetic field, although
this property of the LLL for Dirac fermions is topologically protected for inhomogeneous field
configurations and in the presence of ripples \cite{Giesbers2007PRL}.

The simplest inhomogeneous field configuration which contains nontrivial  Aharonov-Bohm
physics can be created by an infinitesimally thin solenoid. In practice such magnetic field configuration
may be obtained when a type-II superconductor is placed on top of graphene or semicondunducting heterojunction
hosting a 2D electron gas  (2DEG) with quadratic dispersion. While graphene devices still have to be fabricated,
devices like this with a superconducting film grown on top of a semiconducting heterojunction (such as
GaAs/AlGaAs) hosting a 2DEG have in fact been fabricated twenty years ago
\cite{Bending1990PRL,Geim1992PRL} and theoretically well studied
(see e.g. Refs.~\onlinecite{Khaetskii1991JPCM,Brey1993PRB,Moroz1996PRA,Desbois1997NPB,Weeks2007NatPhys}).

Theoretically a problem of the Dirac fermions in the field of Aharonov-Bohm flux was encountered
in the context of cosmic strings by Gerbert and Jackiw \cite{Gerbert1989PRD}. While
for the solutions of the Dirac equation
with nonzero angular momentum \footnote{Notice that the angular momentum operator does not commute with the Dirac Hamiltonian.} the square integrability requirement specifies which of the
two independent solutions should be taken, they noticed this is not the case for the zero angular
momentum.  For the zero momentum there is an ambiguity as both solutions are square integrable,
but divergent as $1/\sqrt{r}$ at the origin, where $r$ is the space coordinate.
The ambiguity of the solution selection is caused by the
singular nature of the infinitesimally thin solenoid vector potential at the origin. This problem
has initiated a vast theoretical literature which addresses interesting aspects related
to the rigorous treatment of the solutions of the Dirac equation with the Aharonov-Bohm potential
(see e.g. Refs.~\onlinecite{Moroz1995PLB,Sitenko2000AP,Gavrilov2003EPJC} for a review).
In the condensed matter context the Dirac fermions in the field of solenoid emerged during the study
of the Dirac-Bogolyubov-de Gennes quasiparticles in the vortex state of $d$-wave superconductors
\cite{Melnikov2001PRL} (see also \cite{Melikyan2007PRB} for a review).
Due to the divergence of the zero mode solutions theory predicts a formation of
nonzero local density of states (LDOS) near the vortex center. However,
this theoretical prediction based on the Dirac nature of quasiparticles in $d$-wave superconductors
does not agree with the results of scanning tunneling spectroscopy (STS) measurements
\cite{Fischer2007RMP} in high-temperature superconductors. Finally we mention a related problem of the description
of topological defects in graphene based on the Dirac equation with a pointlike pseudomagnetic vortex
which has also been studied intensively, see e.g. Refs.~\onlinecite{Vozmediano2010,Sitenko2008NPB}.

The purpose of the present paper is to study the electronic excitations in graphene in the field of
the Aharonov-Bohm flux and compare them with the corresponding results for 2DEG with a
quadratic dispersion. We rely on the existing studies of the Dirac fermions in the
Aharonov-Bohm potential, but focus on the specific features of  graphene such as the presence of
two inequivalent $\mathbf{K}$ points which implies that one should consider the solutions for
both inequivalent irreducible representations of the Dirac $2\times2$ matrices. Also to avoid
unnecessary formal complications we consider the physical regularization of the problem
modeling a finite radius flux tube created by the Abrikosov vortex. We utilize
the simplest case of magnetic field concentrated in a thin cylindrical shell of small but
finite radius $R$ when $R\to 0$ \cite{Alford1989NPB,Hagen1990PRL}.
In contrast to high-temperature superconductors Dirac description of the quasiparticles in graphene
is proven valid under  the different conditions. In particular, STS measurements of
graphene flakes on graphite \cite{Li2009PRL} exhibit the structural and electronic
properties expected of pristine graphene such as the development of a single sequence of
pronounced Landau level peaks corresponding to massless Dirac fermions in  a homogeneous magnetic field.
We propose to perform STS measurements for graphene penetrated by vortices
from a type-II superconductor, because the Dirac theory predicts rather peculiar behavior of LDOS
not expected for the 2DEG with a quadratic dispersion of carriers.

The paper is organized as follows. In Sec.~\ref{sec:model} we introduce the model
Hamiltonians and discuss the regularization of the Aharonov-Bohm potential used in this work.
Sec.~\ref{sec:Schrodinger} is devoted to the nonrelativistic case, and the relativistic case is discussed in detail
in Sec.~\ref{sec:Dirac}. The structure of both sections is the same: we consider the solution of the corresponding
Schr\"{o}dinger or Dirac equation, construct the Green's function (GF) with coinciding arguments, obtain the DOS and study
the behavior of the LDOS.
In Sec.~\ref{sec:concl} our final results are summarized.

\section{Models and main notations}
\label{sec:model}

We consider nonrelativistic and relativistic Hamiltonians.
The 2D nonrelativistic (Schr\"{o}dinger) Hamiltonian has the standard form
\begin{equation}
\label{Hamilton-nonrel}
H_S = -\frac{\hbar^2}{2M} (D_1^2 +D_2^2),
\end{equation}
where $D_j = \nabla_j + i e/\hbar c A_j $, $j=1,2$ with the vector potential $\mathbf{A}$,
Planck's constant $\hbar$ and the velocity of light $c$
describes a spinless particle with a mass $M$ and charge $-e < 0$.
The Dirac quasiparticle in graphene is described by the Hamiltonian
\begin{equation}
\label{Hamilton-rel}
H_D= - i \hbar v_F \beta (\gamma_1 D_1 + \gamma_2 D_2) + \Delta \beta,
\end{equation}
where the matrices $\beta$ and $\beta \gamma_j$ are defined in terms of the Pauli
matrices as
\begin{equation}
\label{matrices}
\beta = \sigma_3, \qquad \beta \gamma_j = (\sigma_1, \zeta \sigma_2).
\end{equation}
Here $\zeta = \pm 1$ labels two unitary inequivalent representations of $2\times2$ gamma matrices,
so that one considers a pair of Dirac equations corresponding to two inequivalent
$\mathbf{K}_{\pm}$ points of graphene's Brillouin zone.
In Eq.~(\ref{Hamilton-rel}) $v_F \approx 10^6 \mbox{m/s}$ is the Fermi velocity and $\Delta$ is the Dirac mass
(gap), which is introduced in the Hamiltonian for generality. Note that we consider the simplest case when the gap has the same
sign for $\zeta = \pm 1$ [see Ref.~\onlinecite{Gusynin2007IJMPB} for a discussion of more general cases].
While tight binding calculations show that the quasiparticle excitations in graphene
have a linear dispersion at low energies \cite{Wallace1947PRev} and are described by the massless Dirac equation
with $\Delta=0$ \cite{Semenoff1984PRL},  recent STS measurements revealed a mass gap near the
Dirac point in a single layer graphene sample suspended above a graphite substrate \cite{Li2009PRL}.
Since this gap and its origin are intensively studied both theoretically and experimentally
in the last few years, here we consider a generic case with a finite value of $\Delta$.

The vector potential of a vortex at the origin directed in the  $\mathbf{e}_z$
direction is
\begin{equation}
\label{A-singular}
\mathbf{A}(\mathbf{r}) = \frac{\Phi}{2 \pi r^2}(\mathbf{r} \times \mathbf{e}_z),
\end{equation}
where $\Phi = \eta \Phi_0$ is the flux of the vortex expressed via magnetic flux quantum
of the electron $\Phi_0 = h c/e$
with $\eta \in [0,1[$, \footnote{In what follows we also consider the behavior of the results
under the reversal of the field direction which corresponds to the negative values of $\eta$.}
where the value $\eta=1/2$ corresponds to the flux created
by the Abrikosov vortex. The magnetic field  is then
\begin{equation}
\mathbf{B}(\mathbf{r}) = \nabla \times \mathbf{A} = \mathbf{e}_z \eta \Phi_0 \delta^2(\mathbf{r}).
\end{equation}

The essential difference between the Schr\"{o}dinger
\begin{equation}
\label{Schrodiger-eq}
H_S \psi = E \psi,
\end{equation}
and Dirac
\begin{equation}
\label{Dirac-eq}
H_D \Psi = E \Psi
\end{equation}
equations in this case can be seen if one squares the latter:
\begin{equation}
\label{Dirac-squared}
-\hbar^2 v_F^2 \left(D_1^2+D_2^2 + i \zeta \sigma_3 [D_1,D_2]  \right) \Psi = (E^2 - \Delta^2) \Psi,
\end{equation}
where the commutator
\begin{equation}
\label{commutator}
i [D_1,D_2] = -\frac{e}{\hbar c} B_z(\mathbf{r})
\end{equation}
which introduces pseudo-Zeeman term which is related to the sublattice rather than
the spin degree of freedom.
It should be mentioned that in the case of graphene the components of the spinor $\Psi(\mathbf{r})$
are associated with a sublattice rather than a spin degree of freedom. Since the Hamiltonian
(\ref{Hamilton-rel}) originates  from a nonrelativistic many-body theory, the Zeeman interaction term
has to be explicitly added to this Hamiltonian. This resembles the case of
the nonrelativistic Hamiltonian (\ref{Hamilton-nonrel}) which becomes Pauli one when the interaction
between the magnetic moment of the spin and an external magnetic field is added. In the present paper
we do not include the spin degree of freedom neither in (\ref{Hamilton-nonrel}) nor in (\ref{Hamilton-rel}).

Eqs.~(\ref{Dirac-squared}) and (\ref{commutator}) identify the origin of complications
\cite{Hagen1990PRL,Jackiw1991book} in the problem with
a singular vortex (\ref{A-singular}) when a singularity in $B_z(\mathbf{r})$ occurs at a singular
point of the differential equation (\ref{Dirac-squared}). To avoid these complications one can
consider a vortex with a finite radius flux tube \cite{Moroz1996PRA,Alford1989NPB,Hagen1990PRL},
i.e. with the magnetic field and vector potential written in cylindric coordinates
$\mathbf{r}=(r,\varphi, z)$:
\begin{equation}
\label{profile}
\mathbf{B}(\mathbf{r}) = \frac{\Phi}{2\pi} h(r) \mathbf{e}_z, \qquad
\mathbf{A}(\mathbf{r}) = \frac{\Phi}{2 \pi} \frac{a(r)}{r} \mathbf{e}_\varphi,
\end{equation}
where $h(r)$ is a profile function with a compact support satisfying the
normalization $\int_0^\infty dr r h(r) =1$ and connected to
the profile function $a(r)$ by the relation $h(r) = a^\prime(r)/r$.

The simplest choice of the field distribution $h(r)$ which regularizes the problem
with the solutions solely expressed in terms of Bessel functions is a magnetic field
concentrated on the surface of the cylinder of radius $R$,
$ h(r)=\delta(r-R)/R$.
Then, the corresponding profile function
\begin{equation}
\label{step-profile}
a(r) = \theta(r-R).
\end{equation}
In the limit $R \to 0$
we recover the Aharonov-Bohm potential (\ref{A-singular}) but avoiding formal complications.
As shown in Ref.~\cite{Hagen1990PRL}, there is no dependence on the detailed form of $h(r)$
in the limit $R\to 0$ provided that $\lim_{r \to 0} \int_0^r h(r^\prime) r^\prime d r^\prime =0$.
In the present paper we restrict ourselves by considering the profile function (\ref{step-profile}).

\section{Nonrelativistic  case}
\label{sec:Schrodinger}

In this section we recapitulate the results of Refs.~\onlinecite{Moroz1996PRA,Desbois1997NPB}
for nonrelativistic case. They are important not only for comparison with relativistic case, but
also because the relativistic result is constructed using the nonrelativistic one.

\subsection{Solutions of the Schr\"{o}dinger equation in Aharonov-Bohm potential and general representation for LDOS}

In the limit $R\rightarrow0$ the admissible solution of the Schr\"{o}dinger equation
is always a regular solution which in polar coordinates
$\mathbf{r}=(r,\varphi)$  takes the form
\begin{equation}
\label{solution-nonrel}
\psi_m(r,\varphi)=\sqrt{\frac{k}{2\pi}}e^{im\varphi}J_{|m+\eta|} (kr), \qquad m \in \mathbb{Z},
\end{equation}
where $J_{|m+\eta|} (kr)$ is the Bessel function with
the wave vector $k$ which is related to the quasiparticle energy $E(k)$ via $E(k) = \hbar^2 k^2/2M$.

The eigenfunction expansion for the retarded Schr\"{o}dinger GF reads
\begin{equation}
\label{nonrel-GF-def}
G_\eta^{\mathrm{S}}(\mathbf{r},\mathbf{r}',E+i0) = \int_0^\infty dk \sum_{m=-\infty}^\infty
\frac{\psi_m (\mathbf{r}) \psi_m^\ast (\mathbf{r}')}{E -E(k)+i0},
\end{equation}
or after substituting the wave function (\ref{solution-nonrel}) it becomes
\begin{equation}
\label{nonrel-GF}
G_\eta^{\mathrm{S}}(\mathbf{r},\mathbf{r}',E+i0) = \frac{M}{\pi\hbar^2} \int_0^\infty
\frac{kdk}{q^2-k^2+i 0} \sum_{m=-\infty}^\infty
e^{im(\varphi-\varphi')} J_{|m+\eta|}(kr)J_{|m+\eta|}(kr'),
\end{equation}
where $q^2=2ME/\hbar^2$. Since an analytic continuation of the GF (\ref{nonrel-GF}) on the imaginary axis
in the complex momentum plane, $q \to z = i\mathcal{Q}$  is free of singularities, it is convenient to work with
the corresponding GF
\begin{equation}
\label{nonrel-GF-continued}
G_\eta^{\mathrm{S}}(\mathbf{r},\mathbf{r}, \mathcal{Q}) \equiv \frac{M}{\pi\hbar^2}
g_\eta (\mathbf{r}, \mathcal{Q}),
\end{equation}
where
\begin{equation}
\label{g-continued}
g_\eta (\mathbf{r}, \mathcal{Q}) = -
\int_0^\infty\
\frac{kdk}{\mathcal{Q}^2+k^2}\sum_{m=-\infty}^\infty J_{|m+\eta|}^2(kr).
\end{equation}
In Eq.~(\ref{nonrel-GF-continued}) we already set two arguments coinciding,
$\mathbf{r} = \mathbf{r}'$, because in the present work we consider the DOS only.
As we will see below, the function $g_\eta (\mathbf{r}, \mathcal{Q})$ is also used in the
representation of the DOS for the Dirac fermions.
After the calculation of the GF (\ref{g-continued}) is done, the LDOS per spin projection
can be found by returning back to the real momentum axis
\begin{equation}
\label{LDOS-def}
N_\eta^{\mathrm{S}}(\mathbf{r},E)=-\frac{1}{\pi}
\mbox{Im}
G_\eta^{\mathrm{S}}(\mathbf{r},\mathbf{r}, \mathcal{Q} \to -i q +  0), \qquad E = \frac{\hbar^2 q^2}{2M}.
\end{equation}
The GF $G_\eta^{\mathrm{S}}(\mathbf{r},\mathbf{r}',\mathcal{Q})$ was calculated in Ref.~\onlinecite{Marino1982NPB}
using the contour integration technique. A weak point of this calculation was discussed
in \cite{Moroz1996PRA}, where the same method was applied to obtain the $\eta$-dependent
contribution to the GF, $\Delta G_\eta^{\mathrm{S}}(\mathbf{r},\mathbf{r},
\mathcal{Q}) = G_\eta^{\mathrm{S}}(\mathbf{r},\mathbf{r}, \mathcal{Q}) -
G_0^{\mathrm{S}}(\mathbf{r},\mathbf{r},\mathcal{Q})$ with coinciding arguments when the approach of \cite{Marino1982NPB}
is valid. Referring to the derivation of  \cite{Marino1982NPB}, here
we simply start from the corresponding expression for $\Delta g_\eta (\mathbf{r},
\mathcal{Q}) =g_\eta (\mathbf{r}, \mathcal{Q})- g_0 (\mathbf{r}, \mathcal{Q}) $ obtained
in Ref.~\onlinecite{Marino1982NPB}
\begin{equation}
\label{Delta-g-Marino}
\Delta g_\eta (\mathbf{r}, \mathcal{Q}) = \frac{\sin \pi \eta}{2\pi}
\int_{-\infty}^\infty dv\int_{-\infty}^\infty d\omega
\frac{e^{\eta(v-\omega)}}{1+e^{(v-\omega)}}e^{-\mathcal{Q}r(\cosh\omega+\cosh v)}.
\end{equation}
Notice that Eq.~(\ref{Delta-g-Marino}) coincides with the corresponding expression from \cite{Moroz1996PRA}
up to a coefficient.
Changing the variables to $x=(v-\omega)/2$, $y=(v+\omega)/2$ one can obtain from (\ref{Delta-g-Marino})
the final expression
\begin{equation}
\label{Delta-g}
\Delta g_\eta (\mathbf{r}, \mathcal{Q})
=\frac{2 \sin\pi\eta}{\pi} \int_0^\infty dy\int_0^\infty dx
\frac{\cosh [(2\eta-1)x]}{\cosh x} e^{-2 \mathcal{Q} r \cosh x\cosh y}
\end{equation}
which we will use in what follows. It turns out that the investigation of the full DOS
is simpler than the analysis of the LDOS. Thus in the next Sec.~\ref{sec:DOS-nonrel}
we firstly consider the DOS and return to the LDOS  (\ref{LDOS-def}) below in Sec.~\ref{sec:LDOS-nonrel}.

\subsection{The density of states}
\label{sec:DOS-nonrel}

The full DOS per spin projection is obtained from the LDOS (\ref{LDOS-def}) by integrating over the space coordinates
\begin{equation}
\label{DOS-nonrel}
N_\eta^{\mathrm{S}}(E)= \int_0^{2\pi} d \varphi \int_0^\infty r d r N_\eta^{\mathrm{S}}(\mathbf{r},E).
\end{equation}
Since we have the integral representation (\ref{Delta-g}) for $\Delta g_\eta (\mathbf{r}, \mathcal{Q})$,
it is straightforward to calculate directly the perturbation of DOS,
$\Delta N_\eta^{\mathrm{S}} (E) =N_\eta (E)- V_{2D} N_0^{\mathrm{S}}$
induced by the Aharonov-Bohm potential. Here $N_0^{\mathrm{S}} = M/(2 \pi \hbar^2)$
is a free DOS of 2DEG per unit area and $V_{2D}$ is the 2D volume (area) of the system.
Firstly integrating
$\Delta g_\eta (\mathbf{r}, \mathcal{Q})$ over space, one obtains \cite{Moroz1996PRA}
\begin{equation}
\int_0^{2\pi} d \varphi \int_0^\infty r d r \Delta g_\eta (\mathbf{r}, \mathcal{Q})
=\frac{\sin\pi\eta}{ \mathcal{Q}^2}\int_0^\infty \frac{dy}{\cosh^2 y}
\int_0^\infty dx\frac{\cosh(2\eta-1)x}{\cosh^3x} .
\end{equation}
Then integrating over $x$ and $y$ we obtain that
\begin{equation}
\label{g-integral}
\int_0^{2\pi} d \varphi \int_0^\infty r d r \Delta g_\eta (\mathbf{r}, \mathcal{Q})
= \frac{\pi \eta (1-\eta)}{ \mathcal{Q}^2} .
\end{equation}
Returning to the real $q$ axis we reproduce the usual Aharonov-Bohm depletion of the DOS
with respect to the free DOS $V_{2D} N_{0}$ \cite{Desbois1997NPB,Moroz1996PRA}:
\begin{equation}
\label{norel-DOS-depletion}
\Delta N_\eta^{\mathrm{S}}(E)=-\frac12 |\eta|(1-|\eta|)\delta(E).
\end{equation}
Writing Eq.~(\ref{norel-DOS-depletion}) we have also included a case of the opposite
field direction.

\subsection{The local density of states}
\label{sec:LDOS-nonrel}

Now we come back to the LDOS (\ref{LDOS-def}). As in the case of the full DOS, it is convenient
to consider the excess LDOS,
$\Delta N_\eta^{\mathrm{S}} (\mathbf{r},E) =N_\eta^{\mathrm{S}} (\mathbf{r},E)- N_0^{\mathrm{S}}$.
Then the value $\Delta N_\eta^{\mathrm{S}} (\mathbf{r},E)$ can be obtained by
calculating the function $\Delta g_\eta (\mathbf{r},\mathcal{Q})$ given by Eq.~(\ref{Delta-g})
and substituting the result to Eq.~(\ref{nonrel-GF-continued}). The analytic continuation $\mathcal{Q}
\to - iq$ described by Eq.~(\ref{LDOS-def}) has to be done at the very last step of the calculation.

Thus our purpose is to derive a simple representation for the function $\Delta g_\eta$.
First, one can rewrite it in the form
\begin{equation}
\label{f-integrand}
\Delta g_\eta(\mathbf{r},\mathcal{Q})=-\int_{\mathcal{Q}}^\infty d w \frac{d \Delta g_\eta(\mathbf{r},w)}{d w},
\end{equation}
where we used that $g_\eta(\mathbf{r},\infty)=0$. Differentiating (\ref{Delta-g}) we get
the integrand of the last expression
\begin{equation}
\frac{d \Delta g_\eta(\mathbf{r},\mathcal{Q})}{d \mathcal{Q}}=
- \frac{4 \sin \pi \eta}{\pi} r\int_0^\infty dx \cosh [(2\eta-1)x]\int_0^\infty dy
 \cosh y e^{-2\mathcal{Q} r\cosh x\cosh y}.
\end{equation}
Using the integral representation for the modified Bessel function $K_\nu(x)$ \cite{Bateman.book2}
\begin{equation}
K_\nu(x)=\int_0^\infty dt e^{-x\cosh t} \cosh \nu t
\end{equation}
and the formula (2.16.13.2) from \cite{Prudnikov.book2}
\begin{equation}
\int_0^\infty dx \cosh b x K_{\nu}( c \cosh x)=\frac12 K_{(\nu+b)/2} \left( \frac{c}{2} \right)K_{(\nu-b)/2}\left( \frac{c}{2} \right),
\end{equation}
we come to the equation
\begin{equation}
\frac{d \Delta g_\eta(\mathbf{r},\mathcal{Q})}{d \mathcal{Q}}=
- \frac{2 \sin \pi \eta}{\pi} r K_\eta(\mathcal{Q} r )K_{1-\eta}(\mathcal{Q} r).
\end{equation}
Integrating the last expression with Mathematica, we get
\begin{equation}
\label{hypergeometric-integral}
\begin{split}
\Delta g_\eta  = &\frac{\sin \pi \eta}{8 \pi}
\Biggl[   4^\eta (\mathcal{Q} r)^{2-2\eta}
\Gamma^2(\eta-1)\,
_2F_3(1-\eta, 3/2-\eta;2-2\eta,2-\eta,2-\eta;\mathcal{Q}^2 r^2) \\
& +  4^{1- \eta} (\mathcal{Q} r)^{2\eta} \Gamma^2(-\eta)\,
_2F_3(\eta, 1/2+\eta;2\eta,1+\eta,1+\eta;\mathcal{Q}^2 r^2) \Biggr] - \ln \frac{\mathcal{Q} r}{2} \\
&  -
\frac{\mathcal{Q}^2 r^2}{4 \eta(1-\eta)}\,_3F_4(1,1,3/2;2,2,2-\eta, 1+\eta;\mathcal{Q}^2 r^2)
 + \frac{1}{2}\psi(1-\eta)+ \frac{1}{2}\psi(\eta) ,
\end{split}
\end{equation}
where $_pF_q(a_1,\ldots a_p;b_1,\ldots b_q;z)$ is the generalized hypergeometric function
and $\psi(z)$ is the logarithmic derivative of the gamma function $\Gamma(z)$.
After the analytic continuation $\mathcal{Q} \to - iq$ is made
only two terms in the square brackets (with the hypergeometric function itself remaining real)
and logarithmic term contribute in $\mbox{Im} \Delta g_\eta(\mathbf{r}, -i q)$, so that
\begin{equation}
\label{LDOS-function}
\Delta N_\eta^{\mathrm{S}}(\mathbf{r},E)= N_0^{\mathrm{S}}\{\sin^2 (\pi \eta)[F(\eta,qr)+F(1-\eta,qr)] -1\}
\end{equation}
with
\begin{equation}
\label{F-eta}
F(\eta,qr)=\frac{4^{\eta-1}(qr)^{2-2\eta}  \Gamma^2(\eta-1)}{\pi^2}
     \,_2F_3 \left(1-\eta,
      \frac32-\eta;2-2\eta,2-\eta,2-\eta;-(qr)^2 \right).
\end{equation}
Note that the last term $-N_0^{\mathrm{S}}$ in Eq.~(\ref{LDOS-function}) arises
from the logarithmic term of (\ref{hypergeometric-integral}).
In the important limiting cases the last expression is greatly simplified.
For  example, in the limit $q r \gg 1$ one obtains
\begin{equation}
\label{LDOS-large}
\Delta N_\eta^{\mathrm{S}}(\mathbf{r},E)= - N_0^{\mathrm{S}}
\frac{\sin (\pi \eta)}{\pi}\frac{\cos(2qr)}{qr}.
\end{equation}
Also in the physically important case $\eta=1/2$ the LDOS is expressed in terms of the sine integral,
$\Si(x) = \int_0^{x} dt \sin t/t$ as follows
\begin{equation}
\label{LDOS-Si}
\Delta N_{1/2}^{\mathrm{S}}(\mathbf{r},E)= N_0^{\mathrm{S}} \left[\frac{2}{\pi} \Si(2 qr)-1 \right].
\end{equation}
Using the asymptotic of the integral sine \cite{Bateman.book2}  $\Si(x) \approx \pi/2 - \cos x/x - \sin x/x^2 + O(1/x^3)$ for $x \gg 1$ from (\ref{LDOS-Si}) we recover the previous expression (\ref{LDOS-large}) valid for $\eta=1/2$ and $q r \gg 1$.
In the opposite limit $\Si(x) =x$ for $x\ll 1$, we see that
$\Delta N_{1/2}^{\mathrm{S}}(\mathbf{r},E)= N_0^{\mathrm{S}}[4 qr/\pi -1]$.

\begin{figure}[h]
\centering{
\includegraphics[width=8cm]{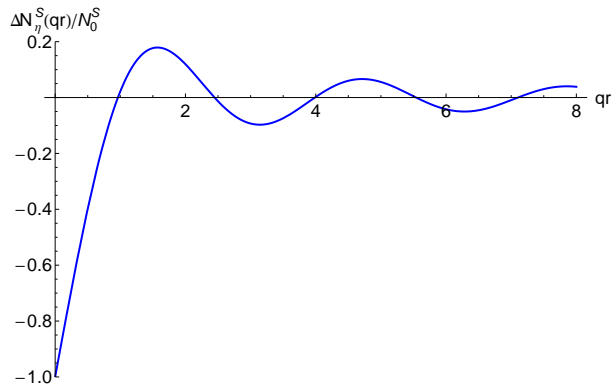}}
\caption{(Color online)
The normalized LDOS function $\Delta N_{\eta}^{\mathrm{S}}(qr)/N_0^{\mathrm{S}}$ as a function of
the dimensionless variable $qr$ for $\eta=1/2$.}
\label{fig:1}
\end{figure}

\begin{figure}[h]
\centering{
\includegraphics[width=8cm]{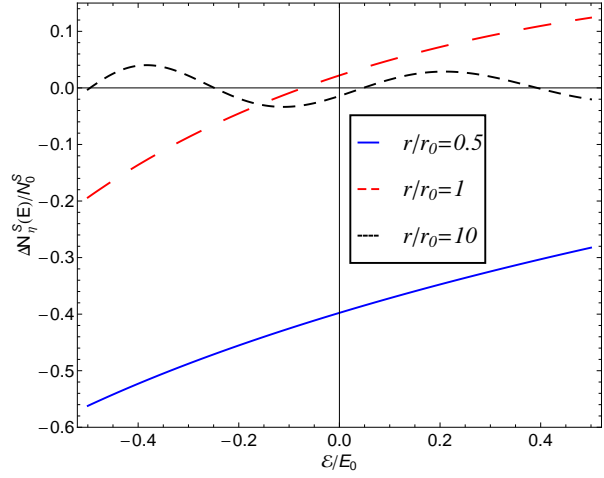}}
\caption{(Color online) The normalized LDOS function $\Delta N_{\eta}^{\mathrm{S}}(\mathcal{E})/N_0^{\mathrm{S}}$ as a function
of energy $\mathcal{E}$ for three values of $r/r_0 = 0.5, 1, 10$ and $\mu = E_0$.}
\label{fig:2}
\end{figure}

In Fig.~\ref{fig:1} we show the dependence (\ref{LDOS-Si}) of the induced LDOS $\Delta N_{1/2}^{\mathrm{S}}(\mathbf{r},E)$ on the distance
from the center of the vortex $r$.
We observe that in the case of nonrelativistic 2DEG the presence of the vortex induced the depletion of the
LDOS for small $q r \ll 1$. The function
$\Delta N_{1/2}^{\mathrm{S}} (qr)$ crosses zero near $q r \approx 1$
and for $q r\approx 1.5$ the function it reaches the maximal value $\sim 0.2 N_{0}^{\mathrm{S}}$.
In Fig.~\ref{fig:2} we model a situation when the STM tip is positioned at some distance from the center of vortex
and a bias voltage is swept to explore the energy dependence of the LDOS. To take into account the presence of a finite
carrier density in 2DEG, we introduce a finite Fermi energy $\mu$, so that the LDOS at zero energy, $\mathcal{E} =0$,
corresponds its value at the Fermi level, i.e. $q = \sqrt{2ME}/\hbar \to q = \sqrt{2M(\mathcal{E}+\mu)}/\hbar$.
To choose the appropriate units we set the distance scale $r_0$ to be the order of the lattice constant.
Then the energy scale, $E_0 = \hbar^2/(2M r_0^2)$ is the order of the bandwidth. The dimensionless variable
$q r$ can now be rewritten as follows $q r = \sqrt{\mathcal{E}/E_0 + \mu/E_0} r/r_0$.
The dependence of  $\Delta N_{\eta}^{\mathrm{S}}(\mathcal{E})$ is shown in Fig.~\ref{fig:2}
for three values of $r/r_0$: for $r/r_0= 0.5$ -- solid (blue) curve, for
$r/r_0 =1$ -- long-dashed (red) curve and for $r/r_0 =10$ -- short-dashed (black) curve. The chemical potential is taken $\mu = E_0$.
We observe that only for the smallest ratio $r/r_0 = 0.5$ the values of the LDOS are
significantly depleted below the free LDOS $N_0^{\mathrm{S}}$. As we saw, the depletion of the LDOS occurs for $ q r \leq 0.5$. Since
the presence of the Fermi surface makes the value of $q$ large, so that the region of small $ qr$ is accessible only for $r \ll r_0$.
Indeed,  we observe that only for the smallest ratio $r/r_0 = 0.5$ the values of the LDOS are
depleted to a half of the value of free LDOS $N_0^{\mathrm{S}}$.
Since the realistic values of the vortex core size are at least of the order of magnitude larger than $r_0$, this implies that
the region of a significant depletion of the LDOS is not accessible experimentally. Still, due to the slow decay of
$\Delta N_{\eta}^{\mathrm{S}} \sim 1/r$ even for $r/r_0 = 10$ the amplitude of $\Delta N_{\eta}^{\mathrm{S}}$ oscillations is of
order of $0.05 N_0^{\mathrm{S}}$, so that this behavior can be probably observed experimentally.

\section{Relativistic  case}
\label{sec:Dirac}
In this section we examine the density of states for Dirac particles in the
potential of the infinitesimally thin solenoid. To avoid formal complications
we consider the physical regularization of the problem with the magnetic field
concentrated in a thin cylindrical shell of small but finite radius $R$ and take the
limit $R \to 0$ at the end of the calculation.
All answers are presented in the form convenient for comparison with the case of the Schr\"{o}dinger equation.

\subsection{Solutions of the Dirac equation in Aharonov-Bohm potential and general representation for LDOS}
\label{sec:Dirac-solutions}

The Dirac equation (\ref{Dirac-eq}) in the field of the regularized vortex (\ref{profile}),
(\ref{step-profile}) (see also (\ref{A-regular})) is solved in Appendix~\ref{sec:A}.
In our consideration we follow Refs.~\onlinecite{Hagen1990PRL,Alford1989NPB}.
The profile (\ref{A-regular}) implies that for $r<R$ the particle obeys
the Dirac equation (\ref{system:r<R}) for a free particle, while for $r>R$
the particle moves in the field of Aharonov-Bohm vortex (\ref{system:r>R}).
Accordingly, for $r<R$ the squared Dirac equation (\ref{Dirac-squared}) is equivalent to
free Schr\"{o}dinger equations for the components of the spinor
\begin{equation}
\Psi(\mathbf{r}) = \left(
                     \begin{array}{c}
                       \psi_1(\mathbf{r}) \\
                       \psi_2(\mathbf{r}) \\
                     \end{array}
                   \right).
\end{equation}
Notice that in the Appendix~\ref{sec:A} the definition (\ref{psi-def-appendix}) for
$\psi_2$ explicitly includes the factor $i$.
For $r>R$ the components $\Psi(\mathbf{r})$ satisfy the Schr\"{o}dinger equations with Aharonov-Bohm
potential and the commutator (\ref{commutator}) is singular at $r=R$:
\begin{equation}
\label{commutator-cylinder}
i[D_1,D_2]=-\frac{\eta}{R}\delta(r-R).
\end{equation}
The solution of the problem can found by matching the solutions
obtained in the domains $r <R $ and $r>R$ [see e.g. (\ref{solution-psi1})]. The radial components of the spinor $\Psi(r)$
have to be continuous:
\begin{equation}
\label{continuity}
\psi_1(R+0)=\psi_1(R-0), \qquad \psi_2(R+0)=\psi_2(R-0),
\end{equation}
and the singularity of the commutator (\ref{commutator-cylinder}) is taken into
account by a condition on the derivatives
\begin{equation}
\label{discontinuity}
\psi_1'(R+0)-\psi_1'(R-0)=\frac{\zeta\eta}{R}\psi_1(R), \qquad \psi_2'(R+0)-\psi_2'(R-0)=-\frac{\zeta\eta}{R}\psi_2(R).
\end{equation}
We stress that in contrast to the Dirac equation case, for the nonrelativistic case
when the solution (\ref{solution-nonrel})  is obtained using the same regularization procedure, both
the wave function and its derivative should be continuous. This as we saw from Eqs.~(\ref{Dirac-squared}),
(\ref{commutator}) and (\ref{components-Cartesian}) is related to the pseudo-Zeeman term.

After the limit $R \to 0$ is taken we obtain for the case $\zeta=1$ the following solutions:
\begin{equation}
\Psi^{(+)}_{m}({\mathbf r})=\sqrt{\frac{k}{4\pi E (k)}}\left(\begin{array}{cc}
 e^{i(m-1)\varphi}\sqrt{E (k)+\Delta}J_{|m+\eta-1|}(kr)\\
\pm i e^{im\varphi}\sqrt{E (k)-\Delta}J_{|m+\eta|}(kr)\end{array}\right),
\end{equation}
for positive value of the energy $E = E(k) = \sqrt{(\hbar v_Fk)^2+\Delta^2}$ and $m\neq0$;
\begin{equation}
\Psi^{(-)}_{m}({\mathbf r})=\sqrt{\frac{k}{4\pi E(k)}}\left(\begin{array}{cc}
e^{i(m-1)\varphi}\sqrt{E(k)-\Delta}J_{|m+\eta-1|}(kr)\\
 \mp i e^{im\varphi}\sqrt{E(k)+\Delta}J_{|m+\eta|}(kr)\end{array}\right),
\end{equation}
for negative value of energy $E = - E(k)$ and $m\neq0$.
Here the upper and lower sign of the second spinor's component corresponds to
$m>0$ and $m<0$ solutions, respectively.
The $m=0$ solution turns out to be special, while the upper component
is regular at $r=0$, the lower component diverges as $J_{-\eta}(kr) \sim r^{-\eta}$:
\begin{equation}
\label{m=0.E>0}
\Psi^{(+)}_{0}({\mathbf r})=\sqrt{\frac{k}{4\pi E(k)}}\left(\begin{array}{cc}
 e^{-i\varphi}\sqrt{E(k)+\Delta}J_{1-\eta}(kr)\\
 -i\sqrt{E(k)-\Delta}J_{-\eta}(kr)\end{array}\right),
\end{equation}
for $E = E(k)$ and
\begin{equation}
\label{m=0.E<0}
\Psi^{(-)}_{0}({\mathbf r})=\sqrt{\frac{k}{4\pi E(k)}}\left(\begin{array}{cc}
e^{-i\varphi}\sqrt{E(k)-\Delta}J_{1-\eta}(kr)\\
 i\sqrt{E(k)+\Delta}J_{-\eta}(kr)\end{array}\right),
\end{equation}
for $E = -E(k)$. One can check this property by using the matching conditions
for $\psi_1(r)$ component and then finding $\psi_2(r)$ from Eq.~(\ref{system:r>R-for2}).
On the other hand, one can discover this singularity in the limit $R\to 0$
by directly analyzing the matching conditions for $\psi_2(r)$ and then finding
nonsingular $\psi_1(r)$ from Eq.~(\ref{system:r>R-for1}).
Comparing Eqs.~(\ref{m=0.E>0}) and (\ref{m=0.E<0}) we observe that a singular at $r=0$
zero-mode solution $E(k) =\Delta$ is hole-like, because a singular electron-like solution
vanishes due to the $\sqrt{E(k) -\Delta}$ factor.
Concluding the discussion of the solutions for the case $\zeta=1$ we stress that
for the opposite field direction it is the upper component of the spinor which is singular in
the solution equivalent to Eqs.~(\ref{m=0.E>0}) and (\ref{m=0.E<0}). Moreover, the nonvanishing zero
mode is now electron-like. As it was firstly noticed in Ref.~\onlinecite{Alford1989NPB}, this
behavior under the change in the field direction breaks the symmetry under $\Phi \to \Phi+\Phi_0$
(see also Ref.~\onlinecite{Beneventano1999IJMPA} for a discussion).

The set of solutions for the case $\zeta=-1$ is the following:
\begin{equation}
\Psi^{(+)}_{m}({\mathbf r})=\sqrt{\frac{k}{4\pi E(k)}}\left(\begin{array}{cc}
 e^{im\varphi}\sqrt{E(k)+\Delta}J_{|m+\eta|}(kr)\\
\mp i e^{i(m-1)\varphi}\sqrt{E(k)-\Delta}J_{|m+\eta-1|}(kr)\end{array}\right),
\end{equation}
for $E = E(k)$ and
\begin{equation}
\Psi^{(-)}_{m}({\mathbf r})=\sqrt{\frac{k}{4\pi E(k)}}\left(\begin{array}{cc}
e^{im\varphi}\sqrt{E(k)-\Delta}J_{|m+\eta|}(kr)\\
 \pm i e^{i(m-1)\varphi}\sqrt{E(k)+\Delta}J_{|m+\eta-1|}(kr)\end{array}\right),
\end{equation}
for $E = -E(k)$. The prescription for
the upper and lower sign of the second spinor's component is the same as for $\zeta=1$.
The $m=0$ solutions in the case $\zeta=-1$ are the following:
\begin{equation}
\label{m=0.E>0-negative-zeta}
\Psi^{(+)}_{0}({\mathbf r})=\sqrt{\frac{k}{4\pi E(k)}}\left(\begin{array}{cc}
 \sqrt{E(k)+\Delta}J_{-\eta}(kr)\\
 ie^{-i\varphi}\sqrt{E(k)-\Delta}J_{1-\eta}(kr)\end{array}\right),
\end{equation}
for $E = E(k)$ and
\begin{equation}
\Psi^{(-)}_{0}({\mathbf r})=\sqrt{\frac{k}{4\pi E(k)}}\left(\begin{array}{cc}
\sqrt{E(k)-\Delta}J_{-\eta}(kr)\\
-ie^{-i\varphi}\sqrt{E(k)+\Delta}J_{1-\eta}(kr)\end{array}\right),
\end{equation}
for $E = -E(k)$. We observe that in this case  it is the upper component of the spinor which is
singular at $r=0$ and nonvanishing is the electron-like zero mode.

Now we are at the position to construct the GF for the Dirac fermions using the presented above solutions.
However, before going to this, we should stress that as shown in\cite{Alford1989NPB,Hagen1990PRL} the
field configuration with $B$ confined to the surface of a cylinder of radius $R$ is not essential
for the main result. In simple words, an more general and not singular at the origin potential can be considered as
a set of concentric shells. Or putting more formally, a profile function $h(r)$ should satisfy the condition
given below Eq.~(\ref{step-profile}) which excludes delta-funtion at the origin.

The eigenfunction expansion for the retarded Dirac GF's now includes both positive and negative
energy solutions
\begin{equation}
G_\eta^{\mathrm{D}}(\mathbf{r},\mathbf{r}',E+i0;\zeta)= \int_0^\infty dk \sum_{m=-\infty}^\infty
\left(\frac{\Psi^{(+)}_{m}(\mathbf{r})\Psi_{m}^{(+)\dag}(\mathbf{r}') }{E-E(k)+i0}+
\frac{\Psi^{(-)}_{m}(\mathbf{r})\Psi_{m}^{(-)\dag}(\mathbf{r}')}{E+E(k)+i0}\right),
\end{equation}
where depending on the sign of $\zeta$ the sum is taken over either $\zeta=1$ or $\zeta=-1$ solutions.
Accordingly, for the diagonal matric elements of $G_\eta^{\mathrm{D}}$, for instance, in the $\zeta=1$ case we obtain
\begin{equation}
\begin{split}
 G^D_{\eta11}(\mathbf{r},\mathbf{r}',E+i0;\zeta=1)  =&\frac{E+\Delta}{2\pi(\hbar v_F)^2}\int_0^\infty \frac{kdk}{q^2-k^2+i 0 \mathrm{sgn} E}\\
& \times \sum_{m=-\infty}^\infty e^{im(\varphi-\varphi')}
J_{|m+\eta|}(kr)J_{|m+\eta|}(kr'),
\end{split}
\end{equation}
and
\begin{equation}
\begin{split}
G^{\mathrm{D}}_{\eta22}(\mathbf{r},\mathbf{r}',E+i0;\zeta=1)=&\frac{E-\Delta}{2\pi(\hbar v_F)^2}\int_0^\infty \frac{kdk}{q^2-k^2+i0 \mathrm{sgn} E}\\
 &\times \sum_{m=-\infty}^\infty e^{im(\varphi-\varphi')}
J_{|m+\eta|}(kr)J_{|m+\eta|}(kr')\\
+&\frac{E-\Delta}{2\pi(\hbar v_F)^2}\int_0^\infty \frac{kdk}{q^2-k^2+i0 \mathrm{sgn} E}[J_{-\eta}(kr)J_{-\eta}(kr')-J_\eta(kr)J_\eta(kr')],
\end{split}
\end{equation}
where $q^2=(E^2-\Delta^2)/(\hbar v_F)^2$.
Then the GF with  coinciding arguments, $\mathbf{r} = \mathbf{r}'$  acquires the following form
\begin{equation}
\begin{split}
\label{Dirac-G-zeta=1}
G^{\mathrm{D}}_{\eta11}(\mathbf{r},\mathbf{r},E+i0;\zeta=1) & =\frac{E+\Delta}{2\pi(\hbar v_F)^2}g_\eta(\mathbf{r},q), \\
G^{\mathrm{D}}_{\eta22}(\mathbf{r},\mathbf{r},E+i0;\zeta=1) & =\frac{E-\Delta}{2\pi(\hbar v_F)^2}[g_\eta(\mathbf{r},q)+ f_\eta(\mathbf{r},q)]
\end{split}
\end{equation}
for $\zeta=1$, and
\begin{equation}
\begin{split}
\label{Dirac-G-zeta=-1}
G^{\mathrm{D}}_{\eta11}(\mathbf{r},\mathbf{r},E+i0;\zeta=-1) & =\frac{E+\Delta}{2\pi(\hbar v_F)^2}[g_\eta(\mathbf{r},q)+f_\eta(\mathbf{r},q)], \\
G^{\mathrm{D}}_{\eta22}(\mathbf{r},\mathbf{r},E+i0;\zeta=-1) & =\frac{E-\Delta}{2\pi(\hbar v_F)^2}g_\eta(\mathbf{r},q),
\end{split}
\end{equation}
for $\zeta=-1$. Here the function $g_\eta(\mathbf{r},q)$ is related to the function $g_\eta (\mathbf{r}, \mathcal{Q})$ defined in Eq.~(\ref{g-continued}) by analytic continuation $q + i0  \mathrm{sgn} E \to z = i  \mathcal{Q} \, \mathrm{sgn} E$.
Note that while the function $g_\eta (\mathbf{r}, \mathcal{Q})$ is identical for both nonrelativistic and relativistic cases,
the real momentum $q$ function $g_\eta(\mathbf{r},q)$  has a different analytical properties in these cases reflecting the fact
that in contrast to the nonrelativistic case the relativistic spectrum contains positive and negative energy branches.
Similarly, the function
\begin{equation}
\label{f-continued}
f_\eta(\mathbf{r}, \mathcal{Q})=-\int_0^\infty \frac{kdk}{\mathcal{Q}^2+k^2}
[J^2_{-\eta}(kr)-J^2_\eta(kr)].
\end{equation}
is the analytic continuation to the imaginary axis of the function $f_\eta(\mathbf{r},q)$.
The $f_\eta$ contribution to $G^{\mathrm{D}}$  originates from the zero mode solutions of the Dirac equation.
Using the integral [see Eq.~(2.12.32.12) from\cite{Prudnikov.book2}]
\begin{equation}
\int_{0}^{\infty} \frac{k d k}{c^2 + k^2} J_\nu^2(kr) = I_\nu (c r) K_\nu (c r),
\end{equation}
where $I_\nu (c r)$ is the modified Bessel function and
$K_\nu (c r)$ is the Macdonald function, we obtain the following simple result for $f_\eta$
\begin{equation}
\label{f-MacDonald}
f_\eta(\mathbf{r}, \mathcal{Q})= - \frac{2 \sin \pi \eta}{\pi} K_\eta^2(\mathcal{Q} r).
\end{equation}

\subsection{The density of states}
\label{sec:DOS-rel}

Due to the matrix structure of the GF and the presence of the valley degree of freedom, $\zeta = \pm1$, in contrast to the
nonrelativistic expression (\ref{DOS-nonrel}), the full DOS in the relativistic case involves not only the integration over area,
but also summing over the diagonal components of the GF and $\mathbf{K}_{\pm}$ valleys.
For a better understanding of the final result for the full DOS, it is instructive to consider separate expressions for
the DOS corresponding separate $\mathbf{K}_{\pm}$ points
\begin{equation}
\label{Dirac-DOS-zeta}
\rho_\eta^{\mathrm{D}}(E,\zeta) = - \frac{1}{\pi} \int_0^{2\pi} d \varphi \int_0^\infty r d r \mbox{Im} \left[\mbox{tr} G^{\mathrm{D}}_{\eta}(\mathbf{r},\mathbf{r},E+i0;\zeta)\right].
\end{equation}
Recall that for $\eta =0$ the free DOS of the Dirac quasiparticles per spin and one valley is equal to
$\rho_0^{\mathrm{D}}(E) = |E| \theta (E^2/\Delta^2 - 1)/(2 \pi \hbar^2 v_F^2) $.

Using Eqs.~(\ref{Dirac-G-zeta=1}) and (\ref{Dirac-G-zeta=-1}) one can express the DOS (\ref{Dirac-DOS-zeta}) via the functions
$g_\eta(\mathbf{r}, \mathcal{Q})$ and $f_\eta(\mathbf{r}, \mathcal{Q})$ defined by Eqs.~(\ref{g-continued}) and (\ref{f-continued}),
respectively, as follows
\begin{equation}
\label{Dirac-DOS-g&f}
\begin{split}
& \rho_\eta^{\mathrm{D}}(E,\zeta) = -\frac{1}{2(\pi\hbar v_F)^2} \\
& \times \int_0^{2\pi} d \varphi \int_0^\infty r d r
\mbox{Im}
\left[2E g_\eta(\mathbf{r}, \mathcal{Q}  \to -i q \,\mathrm{sgn} E +  0 )+
(E-\zeta\Delta)f_\eta(\mathbf{r},\mathcal{Q}  \to -i q \,\mathrm{sgn} E +  0)\right].
\end{split}
\end{equation}
The integral of used in the nonrelativistic case function $g_\eta(\mathbf{r}, \mathcal{Q})$,
or to be precise of the function $\Delta g_\eta$, over the space coordinates is calculated in Eq.~(\ref{g-integral}).
The corresponding space integration of the function $f_\eta(\mathbf{r}, \mathcal{Q})$ given by Eq.~(\ref{f-MacDonald})
produces the result
\begin{equation}
\label{f-integral}
\int_0^{2 \pi} d \varphi \int_0^\infty rdr f_\eta(\mathbf{r},\mathcal{Q})=
-\frac{2\pi\eta}{\mathcal{Q}^{2}}.
\end{equation}
Notice that since $f_\eta =0$ for $\eta =0$, there is no need to introduce a function $\Delta f_\eta$.
Having these results for the spatial integration we can calculate the imaginary part.
Since both Eqs.~(\ref{g-integral}) and (\ref{f-integral}) depend on $\mathcal{Q}^2$, the imaginary part is obtained by using
a simple prescription $\mathcal{Q}^2 \to - q^2 = -[(E+i0)^2-\Delta^2]/(\hbar v_F)^2$. This gives
the final expression for the perturbed DOS,
$\Delta \rho_\eta^{\mathrm{D}}(E,\zeta) = \rho_\eta^{\mathrm{D}}(E,\zeta) - V_{2D} \rho_0^{\mathrm{D}}(E) $ by the
Aharonov-Bohm vortex
\begin{equation}
\label{Dirac-DOS-final}
\Delta \rho_\eta^{\mathrm{D}} (E,\zeta)=-\frac{1}{2}|\eta|(1-|\eta|)[\delta(E-\Delta)+\delta(E+\Delta)]+
|\eta| \delta (E + \zeta \mbox{sgn} (\eta) \Delta).
\end{equation}
The first term of (\ref{Dirac-DOS-final}) which is $\sim -|\eta| (1-|\eta|)$ originates from  $g_\eta$ part of
Eq.~(\ref{Dirac-DOS-g&f}) and resembles the nonrelativistic result (\ref{norel-DOS-depletion}). For $\Delta=0$ it
turns out to be twice larger than (\ref{norel-DOS-depletion}) simply because in Eq.~(\ref{Dirac-DOS-final}) we summed over
the diagonal components of the GF $G_\eta^{\mathrm{D}}$ which are related to the two sublattices of graphene.
The last term of (\ref{Dirac-DOS-final}) originates from  $f_\eta$ part of Eq.~(\ref{Dirac-DOS-g&f}) and thus is related to the zero
mode solution of the Dirac equation.  The mentioned above fact that the singular component of the zero mode solution
is hole-like for $\zeta=1$ [see Eq.~(\ref{m=0.E<0})] and it is electron-like for $\zeta=-1$ [see Eq.~(\ref{m=0.E>0-negative-zeta})]
finds its reflection in the asymmetric form of the last term of Eq.~(\ref{Dirac-DOS-final}) which  also corresponds to
the holes (electrons) for $\zeta =1$ ($\zeta=-1$). In accordance with the behavior of the zero mode solutions described in Sec.~\ref{sec:Dirac-solutions}, when the direction of the field is reversed the expressions $\rho_\eta (E,\zeta=1)$ and
$\rho_\eta (E,\zeta=-1)$ are interchanged. The full excess DOS
\begin{equation}
\label{Dirac-DOS-full}
\Delta N_\eta^{\mathrm{D}} (E) = \Delta \rho_\eta^{\mathrm{D}} (E,\zeta=1) + \Delta \rho_\eta^{\mathrm{D}} (E,\zeta=-1) =
\eta^2 [\delta(E-\Delta)+\delta(E+\Delta)]
\end{equation}
is obviously symmetric in energy. In contrast to the nonrelativistic case, in the Dirac case the Aharonov-Bohm vortex induces the
excess of the states which is related to the presence of the last term of Eq.~(\ref{Dirac-DOS-final}) and caused by the zero modes.
We note that in Ref.~\onlinecite{Moroz1995PLB} the corresponding term of Eq.~(\ref{Dirac-DOS-final}) has a wrong sign.
The positiveness of $\Delta N_\eta^{\mathrm{D}} (E)$ can also be understood by the following simple argument. For $\Delta =0 $
the free DOS $\rho_0^{\mathrm{D}}(E=0)=0$. Therefore, since the DOS has to be positive, the value $\Delta N_\eta^{\mathrm{D}} (E)$
should also be positive.

\subsection{The local density of states}

Now we investigate the LDOS for the Dirac case. While it was useful to consider each valley separately, especially because
field theoretical studies of the problem often involve only one unitary inequivalent representation of $2 \times 2$ gamma matrices
[see e.g.\cite{Moroz1995PLB}], the LDOS measurement picks up both valleys together. On the other hand, LDOS distinguishes
sublattices.  Thus we consider separately the LDOS for $A$ and $B$ sublattices which are defined as follows
\begin{equation}
\label{LDOS-AB}
\begin{split}
N_\eta^{\mathrm{D} (A)}(\mathbf{r},E)=-\frac{1}{\pi} \mbox{Im}
\left[ G_{\eta 11}(\mathbf{r},\mathbf{r},E+i0;\zeta=1) + G_{\eta11}(\mathbf{r},\mathbf{r},E+i0;\zeta=-1) \right], \\
N_\eta^{\mathrm{D} (B)}(\mathbf{r},E)=-\frac{1}{\pi} \mbox{Im}
\left[ G_{\eta 22}(\mathbf{r},\mathbf{r},E+i0;\zeta=1) + G_{\eta22}(\mathbf{r},\mathbf{r},E+i0;\zeta=-1) \right].
\end{split}
\end{equation}
Again we consider the perturbation of the LDOS induced by the Aharonov-Bohm potential,
$\Delta N_\eta^{\mathrm{D} (A,B)}(\mathbf{r},E)=N_\eta^{\mathrm{D} (A,B)}(\mathbf{r},E)-N_0^{\mathrm{D} (A,B)}(\mathbf{r},E)$.
Using Eqs.~(\ref{Dirac-G-zeta=1}) and (\ref{Dirac-G-zeta=-1}) we rewrite the LDOS's
$\Delta N_\eta^{\mathrm{D} (A,B)}(\mathbf{r},E)$ in terms of the functions
$\Delta g_\eta(\mathbf{r}, \mathcal{Q})$ and $f_\eta(\mathbf{r}, \mathcal{Q})$ as follows
\begin{equation}
\label{LDOS-AB-f&g}
\Delta N_\eta^{\mathrm{D} (A,B)}(\mathbf{r},E)=-\frac{E\pm\Delta}{2(\pi\hbar v_F)^2}
\mbox{Im}[2 \Delta g_\eta(\mathbf{r},\mathcal{Q}  \to -i q \,\mathrm{sgn} E +  0)+
f_\eta(\mathbf{r},\mathcal{Q}  \to -i q \,\mathrm{sgn} E +  0)].
\end{equation}
Here the upper (lower) sign corresponds to $A$ ($B$) sublattice. To obtain the final expression for LDOS we should
make the analytic continuation  $\mathcal{Q}  \to -i q \, \mbox{sgn} E$.
For the function $\Delta g_\eta(\mathbf{r},\mathcal{Q})$ given by Eq.~(\ref{hypergeometric-integral})
one obtains
\begin{equation}
\label{g-real-rel}
\mbox{Im} \Delta g_\eta(\mathbf{r},\mathcal{Q}  \to -i q \, \mbox{sgn} E)=-\frac{\pi \mathrm{sgn} E}{2}\left\{
\sin^2\pi\eta[F(\eta,qr)+F(1-\eta,qr)]-1\right\},
\end{equation}
where the function $F(\eta,qr)$ is defined in Eq.~(\ref{F-eta}). One can easily see that because
in the nonrelativistic case the analytic continuation to  the real momentum $\mathcal{Q} \to - iq$
is different from the analytic continuation  $\mathcal{Q}  \to -i q \, \mbox{sgn} E$
in the relativistic case, the function (\ref{g-real-rel}) differs from the function  (\ref{LDOS-function}).
Using the relationships \cite{Bateman.book2}
\begin{equation}
K_\nu(z e^{i \frac{\pi}{2}}) = - i \frac{\pi}{2} e^{- i\frac{\pi \nu}{2}} H^{(2)}_\nu (z), \qquad H^{(2)}_\nu (z) = J_{\nu}(z) - i Y_\nu(z)
\end{equation}
between the Macdonald function of the imaginary argument, the Hankel function of the second kind $H^{(2)}_\nu(z)$ and the Bessel function of the
first $J_\nu(z)$ and second  $Y_\nu(z)$ kinds, we obtain that
\begin{equation}
K^2_\eta(\pm iz)=\frac{\pi^2}{4}e^{ \mp i\pi\eta}[Y^2_\eta(z)-J^2_\eta(z) \pm 2i J_\eta(z)Y_\eta(z)],
\end{equation}
where we used the property $K_\nu(z^\ast) = K_\nu^\ast(z)$.
Accordingly,  the analytic continuation of the function $f_\eta(\mathbf{r},\mathcal{Q})$ from
Eq.~(\ref{f-MacDonald}) takes the form
\begin{equation}
\label{f-real-rel}
\mbox{Im} f_\eta(\mathbf{r},\mathcal{Q}  \to -i q \, \mbox{sgn} E)= - \frac{\pi\sin\pi\eta \, \mathrm{sgn} E}{2}
\{\sin\pi\eta[Y^2_\eta(qr) - J^2_\eta(qr)]-2\cos\pi\eta J_\eta(qr)Y_\eta(qr)\}.
\end{equation}
Substituting (\ref{g-real-rel}) and (\ref{f-real-rel}) in Eq.~(\ref{LDOS-AB-f&g}) we arrive at the final main result
\begin{equation}
\begin{split}
\label{LDOS-AB-final}
\Delta N_\eta^{\mathrm{D} (A,B)}(\mathbf{r},E)= N_0^{\mathrm{D}} (E) & \left(1 \pm\frac\Delta E\right) \theta \left(\frac{E^2}{\Delta^2}-1 \right)
\Biggl\{ \sin^2(\pi\eta)[F(\eta,qr)+F(1-\eta,qr)]-1\\
+& \frac{\sin^2(\pi\eta)}{2} [Y^2_\eta(qr)- J^2_\eta(qr)]- \frac{\sin( 2 \pi\eta)}{2} J_\eta(qr)Y_\eta(qr) \Biggr\}, \qquad
q = \frac{\sqrt{E^2 - \Delta^2}}{\hbar v_F},
\end{split}
\end{equation}
where $N_0^{\mathrm{D}} (E)=|E|/(2 \pi\hbar^2 v_F^2)$ is free DOS of the Dirac quasiparticles per spin  and one sublattice (or valley) for $\Delta=0$.
The first part of Eq.~(\ref{LDOS-AB-final}) which includes $F$ and $-1$ is identical to the nonrelativistic expression (\ref{LDOS-function}),
while the second part of Eq.~(\ref{LDOS-AB-final}) with Bessel functions originates from the zero mode contribution.

In the limit $q r \gg 1$ the last expression acquires a simple form
\begin{equation}
\label{LDOS-AB-large-qr}
\Delta N_\eta^{\mathrm{D} (A,B)}(\mathbf{r},E)= - N_0^{\mathrm{D}} (E) \left(1 \pm\frac\Delta E\right) \theta \left(\frac{E^2}{\Delta^2}-1 \right)
\frac{\eta \sin (\pi \eta)}{\pi} \frac{\sin ( 2 q r)}{q^2 r^2}.
\end{equation}
Comparing Eqs.~(\ref{LDOS-AB-large-qr}) and (\ref{LDOS-large}) we conclude that in the Dirac case the impact of the vortex is
more localized than in the nonrelativistic case.

As we saw in Sec.~\ref{sec:LDOS-nonrel} in the physically important case $\eta=1/2$ the expression (\ref{LDOS-function}) is significantly
simplified to the result (\ref{LDOS-Si}). The same remains true for Eq.~(\ref{LDOS-AB-final}), because half-integer Bessel functions
are expressed in terms of the elementary functions and we obtain that
\begin{equation}
\label{LDOS-AB-Si-cos}
\Delta N_\eta^{\mathrm{D} (A,B)}(\mathbf{r},E)= N_0^{\mathrm{D}} (E) \left(1 \pm\frac\Delta E\right)\theta \left(\frac{E^2}{\Delta^2}-1 \right)\left[\frac{2}{\pi} \Si(2 qr)-1 +
\frac{\cos( 2 qr)}{\pi qr}\right].
\end{equation}
From Eq.~(\ref{LDOS-AB-Si-cos}) we immediately observe the main difference between the relativistic and nonrelativistic cases.
The presence of zero modes causes a positive divergence of the LDOS,  $\Delta N_\eta^{\mathrm{D} (A,B)}(\mathbf{r},E)\sim 1/qr$ for $q r \ll 1$
near the center of the vortex. Integrally this results in the
excess of the states in the full DOS (\ref{Dirac-DOS-full}).
Using the asymptotic expansion of $\Si(x)$ given below Eq.~(\ref{LDOS-Si})
we recover the previous expression (\ref{LDOS-AB-large-qr}) valid for $\eta=1/2$ and $q r \gg 1$.

\begin{figure}[h]
\centering{
\includegraphics[width=8cm]{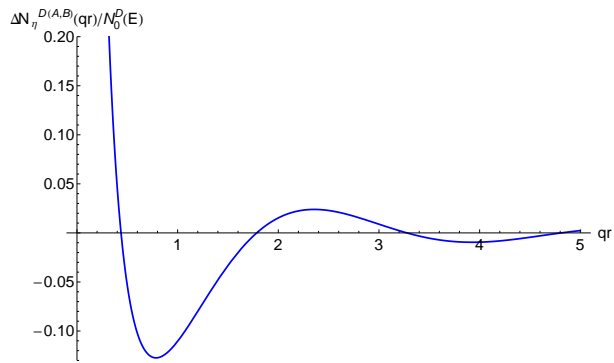}}
\caption{(Color online) The normalized LDOS function $\Delta N_{\eta}^{\mathrm{D}(A,B)}(qr)/N_0^{\mathrm{D}}(E)$ as a function of
the dimensionless variable $qr$ for $\eta=1/2$.}
\label{fig:3}
\end{figure}

In Fig.~\ref{fig:3} we show the dependence (\ref{LDOS-AB-Si-cos}) of the induced LDOS
$\Delta N_{1/2}^{\mathrm{D}(A,B)}(\mathbf{r},E)$ on the distance from the center of the vortex $r$ for $\Delta=0$.
We observe the features expected from the analytic expressions such as the excess of the LDOS for small $q r \ll 1$ and
faster than in 2DEG decay of $\Delta N_{1/2}^{\mathrm{D}} \sim 1/r^2$ for $q r \gg 1$.

\begin{figure}[h]
\centering{
\includegraphics[width=8cm]{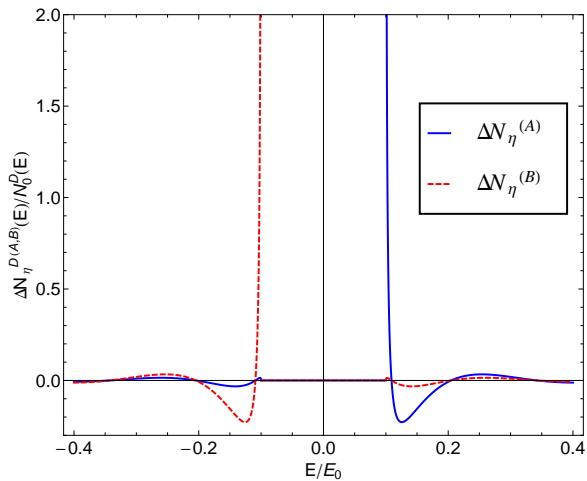}}
\caption{(Color online) The normalized LDOS function $\Delta N_{\eta}^{\mathrm{D}(A,B)}(E)/N_0^{\mathrm{D}}(E)$
as a function of energy $E$ for $r/r_0 = 10$, $\Delta =0.1 E_0$ and $\eta=1/2$.}
\label{fig:4}
\end{figure}

In Fig.~\ref{fig:4} we consider a situation similar to Fig.~\ref{fig:2}. We fix the distance at $r = 1 0 r_0$ and plot the energy
dependence of the relativistic LDOS (\ref{LDOS-AB-Si-cos}). We consider the most interesting case of undoped graphene
with zero carrier density. In contrast to the 2DEG, in graphene is easily tuned to this regime.
Again, we introduce  the distance scale $r_0$ of the order of the lattice constant.
Then for the Dirac case the energy scale is $E_0=\hbar v_F/r_0$, and accordingly
the dimensionless variable is $q r = \sqrt{(E/E_0)^2-(\Delta/E_0)^2}r/r_0$.
To make sublattices inequivalent we introduce a finite gap $\Delta = 0.1 E_0$ which introduces the asymmetry between the
the LDOS on $A$ and $B$ sublattices.
The LDOS $\Delta N_{1/2}^{\mathrm{D}(A)}(E)$ shown as a solid (blue) curve has a sharp peak near $E = \Delta$ and practically
no peak at $E = \Delta$, while the  LDOS $\Delta N_{1/2}^{\mathrm{D}(B)}(E)$ shown as dashed (red) curve
has a peak near $E = - \Delta$ and no peak at $E =  \Delta$.
When $E$ is increasing the LDOS very quickly reduces to its value for the Dirac system in the absence of the vortex.
We note that this behavior of the LDOS is seen for the large value of $r/r_0 = 10$, i.e. far from the center of the vortex,
indicates that it should be possible to observe this excess of the LDOS in experiments
on graphene.

Finishing our discussion of the Dirac case, we mention a recent work \cite{Jackiw2009PRB}, where the effect of the vacuum polarization
in the field of an infinitesimally thin solenoid at the distances much larger than the radius of solenoid was studied.
Constructing the GF the authors neglected the delta-function (\ref{discontinuity}) motivated by the fact that they are interested
in the regime $r \gg R$. One of their interesting conclusions is that for $\eta=1/2$ the induced current is zero.

\section{Conclusions}
\label{sec:concl}

The main motivation of this work was to address a natural question whether one can distinguish
graphene from 2DEG measuring the LDOS near the Abrikosov vortex penetrating them. Just by comparing Figs.~\ref{fig:2}
and \ref{fig:4} we find a positive answer. Indeed, by putting the STM tip close to the vortex, for example,
at the distance, $r \sim 10 r_0$, where $r_0$ is the lattice constant, we observe that in 2DEG the value of the LDOS will
be close to its background value $N_0^\mathrm{S}$ in the free system, while in undoped graphene one should observe a
strong LDOS enhancement near the Fermi level. Or to be more precise, if there is no gap $\Delta$ in the quasiparticle spectrum,
the peak of the LDOS should be observed at the Fermi level, or if $\Delta\neq 0$ the peaks should be observed at the energies
$E = \pm \Delta$ and the sign of the peak energy should depend on the sublattice.
This peaked behavior of the LDOS with $\Delta N_\eta^{\mathrm{D} (A,B)}(\mathbf{r},E)\sim 1/qr$ for $q r \ll 1$
reflects the specific feature of the Dirac fermions such as the presence of the divergent as
$1/\sqrt{r}$ at the origin zero mode solution of the Dirac equation. Thus the observation of this feature
in STS measurements would contribute to the expanding list of the experimental manifestations of the Dirac fermions
in graphene. Our second conclusion is that while in the nonrelativistic case the presence of the Aharonov-Bohm vortex leads to a depletion
of the full DOS, in the Dirac case the full DOS is enhanced. This result can likely be checked by analyzing the STS maps. We remind at the
end that in this paper we did not consider the effect of disorder in the presence of magnetic field
which will definitely affect the behavior of the LDOS in magnetic field \cite{Champel2010} and should be
taken into account in the  analysis of the STS data.

\section{Acknowledgments}

We thank E.V.~Gorbar and V.P.~Gusynin for many stimulating discussions.
S.G.S. thanks C.G.~Beneventano and E.M.~Santangelo for an interesting discussion and
bringing Ref.~\onlinecite{Beneventano1999IJMPA} to his attention.
This work was supported by the SCOPES grant No. IZ73Z0\verb|_|128026 of
Swiss NSF, by the grant SIMTECH No. 246937 of the European FP7 program and by
the Program of Fundamental Research of the Physics and Astronomy Division of
the National Academy of Sciences of Ukraine.
S.G.S. was also supported by
the Ukrainian State Foundation for Fundamental
Research under Grant No. F28.2/083.

\appendix
\section{Solution of the Dirac equation}
\label{sec:A}

A positive energy solution of the time-dependent Dirac equation has a form
$\Psi(t, \mathbf{r}) = \exp(-i Et/\hbar)\Psi(\mathbf{r})$ with
a two-component spinor $\Psi(\mathbf{r})$
satisfying  the time-independent Dirac equation (\ref{Dirac-eq}):
\begin{equation}
\label{Dirac-eq-append}
\left[i \hbar v_F \sigma_1 \left(\partial_1 + i \frac{e}{\hbar c} A_1 \right)+
      i \hbar v_F \zeta \sigma_2 \left(\partial_2 + i \frac{e}{\hbar c} A_2 \right)
- \Delta \sigma_3 \right]\Psi({\mathbf r})=0.
\end{equation}
where $\zeta = \pm 1$ distinguishes two unitary inequivalent representations of $2\times 2$
gamma matrices.
It is convenient to denote the components of two-component spinor $\Psi(\mathbf{r})$
\begin{equation}
\label{psi-def-appendix}
\Psi(\mathbf{r}) = \left(\begin{array}{cc} \psi_1({\mathbf r})\\
i\psi_2({\mathbf r})  \end{array}\right)
\end{equation}
with the factor  $i$ explicitly included in the definition of  the lower component.
Then  we can rewrite Dirac equation (\ref{Dirac-eq-append}) in the components
as follows:
\begin{equation}
\label{components-Cartesian}
\begin{split}
(E-\Delta)\psi_1({\mathbf r})-\hbar v_F(D_1-i\zeta D_2)\psi_2({\mathbf r})= &0,\\
\hbar v_F(D_1+i\zeta D_2)\psi_1({\mathbf r})+(E+\Delta)\psi_2({\mathbf r})= &0.
\end{split}
\end{equation}
Since we consider a cylindrically symmetric configuration of the field with a vector
potential $\mathbf{A}=\mathbf{e}_\varphi A_\varphi(r)$, the system (\ref{components-Cartesian})
has to be rewritten in the polar coordinates $(r,\varphi)$:
\begin{equation}
\label{components-polar}
\begin{split}
(E-\Delta)\psi_1(\mathbf{r})-\hbar v_Fe^{-i\zeta\varphi}\left(\frac{\partial}
{\partial r}-\frac{i\zeta}{r}\frac{\partial}{\partial\varphi}+
\frac{e\zeta A_\varphi}{\hbar c}\right)\psi_2(\mathbf{r})= & 0 , \\
\hbar v_Fe^{i\zeta\varphi}\left(\frac{\partial}
{\partial r}+\frac{i\zeta}{r}\frac{\partial}{\partial\varphi}-
\frac{e\zeta A_\varphi}{\hbar c}\right)\psi_1(\mathbf{r})
+(E+\Delta)\psi_2(\mathbf{r})= & 0.
\end{split}
\end{equation}
As discussed in Sec.~\ref{sec:model}, to analyze the problem with a singular at $r=0$ Aharonov-Bohm potential (\ref{A-singular}) one has to do a self-adjoint extension of the Dirac operator, see e.g.
Refs.~\onlinecite{Gerbert1989PRD,Sitenko2000AP,Jackiw1991book}. To avoid this complication we consider a regularized
field configuration  (\ref{profile})  suggested in Refs.~\onlinecite{Alford1989NPB,Hagen1990PRL}
with the profile function (\ref{step-profile}), so that
\begin{equation}\label{A-regular}
\frac{e A_\varphi(r)}{\hbar c}=\begin{cases}
             0, \qquad r<R,\\
             \eta/r, \quad r>R.
             \end{cases}
\end{equation}
From now on, we consider the specific case $\zeta =1$ and
seek for a solution of Eq.~(\ref{components-polar}) in the following form
\begin{equation}
\psi_1(\mathbf{r})=e^{i(m-1)\varphi}\psi_1(r),\qquad \psi_2(\mathbf{r})=e^{im\varphi}\psi_2(r)
\end{equation}
for $r < R$  we obtain a system of the radial equations for a free Dirac particle
\begin{subequations}
\label{system:r<R}
\begin{align}
(E-\Delta)\psi_1(r) -\hbar v_F \left[\frac{d}{dr}+\frac{m}{r}\right]\psi_2(r)= & 0, \\
\hbar v_F\left[\frac{d}{dr}-\frac{m-1}{r}\right]\psi_1(r) +(E+\Delta)\psi_2(r)= &0,
\end{align}
\end{subequations}
while for $r > R$ we have
\begin{subequations}
\label{system:r>R}
\begin{align}
(E-\Delta)\psi_1(r)-\hbar v_F\left[\frac{d} {dr}+\frac{1}{r}(m+\eta)\right]\psi_2(r)= & 0,  \label{system:r>R-for1} \\
\hbar v_F\left[\frac{d}{dr}-\frac{1}{r}(m+\eta-1)\right]\psi_1(r)+(E+\Delta)\psi_2(r)= &0.  \label{system:r>R-for2}
\end{align}
\end{subequations}
One can obtain from the systems (\ref{system:r<R}) and (\ref{system:r>R}) that the spinor components
satisfy the following second order differential
equations:
\begin{subequations}
\label{2nd-order:r<R}
\begin{align}
& \frac{d^2}{dr^2}\psi_1(r)+\frac{1}{r}\frac{d}{dr}\psi_1(r)-
\left[\frac{(m-1)^2}{r^2}-\frac{E^2-\Delta^2}{(\hbar v_F)^2}\right]\psi_1(r)=0,
\label{psi1:r<R} \\
& \frac{d^2}{dr^2}\psi_2(r)+\frac{1}{r}\frac{d}{dr}\psi_2(r)-
\left[\frac{m^2}{r^2}-\frac{E^2-\Delta^2}{(\hbar v_F)^2}\right]\psi_2(r)=0, \label{psi2:r<R}
\end{align}
\end{subequations}
for $r < R$ and
\begin{subequations}
\label{2nd-order:r>R}
\begin{align}
\frac{d^2}{dr^2}\psi_1(r)+\frac{1}{r}\frac{d}{dr}\psi_1(r)-
\left[\frac{(m+\eta-1)^2}{r^2}-\frac{E^2-\Delta^2}{(\hbar v_F)^2}\right]\psi_1(r)=0,
\label{psi1:r>R} \\
\frac{d^2}{dr^2}\psi_2(r)+\frac{1}{r}\frac{d}{dr}\psi_2(r)-
\left[\frac{(m+\eta)^2}{r^2}-\frac{E^2-\Delta^2}{(\hbar v_F)^2}\right]\psi_2(r)=0 \label{psi2:r>R}
\end{align}
\end{subequations}
for $r>R$. The solutions of Eqs.~(\ref{2nd-order:r<R}) and (\ref{2nd-order:r>R}) are expressed in terms of
the usual  Bessel functions. For example, for the solutions of the equations (\ref{psi1:r<R}) and
(\ref{psi1:r>R}) for the component $\psi_1(r)$ are given by
\begin{subequations}
\label{solution-psi1}
\begin{align}
\psi_1(r)& =C_m J_{|m-1|}(kr), \qquad r<R,
\label{solution-psi1:r<R} \\
\psi_1(r)&=A_m J_{|m+\eta-1|}(kr)+ B_m J_{-|m+\eta-1|}(kr), \qquad r>R, \label{solution-psi1:r>R}
\end{align}
\end{subequations}
where $A_m$, $B_m$, and $C_m$ are constants and the $J$'s are the Bessel functions.
The solution (\ref{solution-psi1:r<R}) is standard due to normalizability and the absence
of delta function at $r=0$. The coefficients $A_m$, $B_m$, and $C_m$ to be found from the matching conditions (\ref{continuity}) and (\ref{discontinuity}). To find the second component, $\psi_2(r)$
one can substitute the result for  $\psi_1(r)$ in Eq.~(\ref{system:r>R-for2}).
Or, equivalently, one can start from Eqs.~(\ref{psi2:r<R}) and
(\ref{psi2:r>R}) for the component $\psi_2(r)$ which also have a solution in the form (\ref{solution-psi1}),
find the corresponding constants from the matching conditions and then use Eq.~(\ref{system:r>R-for1}) to obtain  $\psi_1(r)$.
Finally, the overall factor before
the solution is determined by the normalization condition
\begin{equation}
\int_{0}^{2 \pi} d \varphi \int_{0}^{\infty} r d r \Psi^\dagger_{m^\prime}(r,\varphi;k^\prime)
\Psi_m(r,\varphi;k) = \delta(k-k^\prime) \delta_{m,m^\prime}.
\end{equation}

\end{document}